\documentclass[12pt]{article}
\usepackage{amscd, amssymb}
\usepackage{multirow}
\usepackage[mathscr]{eucal}
\newtheorem{thm}{Theorem}[section]

\newtheorem{ex}[thm]{Example}
\newtheorem{defn}[thm]{Definition}

\newcommand{\eep}{\hfill $\square$}
\newcommand{\pf}{\noindent {\bf Proof. \ }}
\usepackage{epsfig}
\title{Double Cylinder Cycle codes of Arbitrary Girth} \author{Mehdi Samadieh\thanks{Corresponding author: Mehdi Samadieh is the Researcher of Isfahan Mathematics House, Isfahan, IRAN, e-mail: samadieh.m@gmail.com, m.samadieh@mathhouse.org} and
 \newline
 ~Mohammad Gholami\thanks{
Mohammad Gholami is with the Department of Mathematics, Shahrekord
University,
 115, Shahrekord, IRAN, e-mail: gholami-m@sci.sku.ac.ir, gholamimoh@gmail.com.}}
= 6.7in
\begin{document}
\baselineskip=25pt
\maketitle
\begin{abstract}
\footnotesize A particular class of low-density parity-check codes referred to as cylinder-type BC-LDPC codes is proposed by Gholami and Eesmaeili. In this paper We represent a double cylinder-type parity-check matrix H by a graph called the block-structure graph of H and denoted by BSG(H). Using the properties of BSG(H) we propose some mother matrices with column-weight two such that the rate of corresponding cycle codes are greater tan cycle codes constructed by Gholami with same girth.\vspace{.2cm}
\end{abstract}
{\bf Keywords:} LDPC codes, tanner graph, girth, closed walk.
\section{Introduction}
Low-density parity-check (LDPC) codes are a class of linear block
codes represented by sparse parity-check matrices~\cite{gal},
capable of performing very close to the Shannon capacity limits
when they are decoded under simple iterative
decoders~\cite{mac1},~\cite{richard}, such as Sum-product
algorithm~\cite{sum}. LDPC codes can be constructed into two
methods: random codes~\cite{mac1}-\cite{bone} and structured
codes~\cite{tan1}-\cite{sun}. Quasi-cyclic(QC) LDPC codes are the most
promising class of structured LDPC codes due to their ease of
implementation and excellent performance over noisy channels when
decoded with message-passing algorithms as extensive simulation
studies have shown. Compared with randomly constructed codes,
QC-LDPC codes can be encoded in linear time with shift registers
and require small memory space to store the code graphs for
decoding. LDPC codes are specified by a sparse parity check matrix
and its corresponding Tanner graph~\cite{tan1}. Tanner determined
a lower bound on the minimum distance that grows exponentially
with the girth of the code. The extension of LDPC codes to
non-binary Galois field GF(q) was first investigated empirically
by Davey and MacKay over the binary-input AWGN
channel~\cite{mac2}.
\medskip

Cycle codes~\cite{cyclecodes} are a special class of binary LDPC
codes with the property that each column of the parity-check
matrix contains exactly two nonzero elements. Gallager [1] has
shown that the minimum distance of cycle codes increases
logarithmically with code length, while this increment is linear
for the codes with column-weight greater than two. In spite of
this weakness which causes to have poor performance, cycle codes
have their own advantages: 1) Their encoding and decoding
operations have lower computation and storage complexity, 2) They
have better block error statistics when applied on partial
response channels~\cite{song1}-\cite{song3}, 3) Non-binary cycle
codes are among the most promising non-binary codes and for q =
64, 128, 256, the best q-ary LDPC codes decoded with belief
propagation (BP) algorithm are cycle
codes~\cite{cycle1}-\cite{cycle2}. This makes non-binary cycle
codes good candidates for both optimum maximum likelihood (ML) and
iterative decoding, 4) Compared with other LDPC codes, the girth
of the Tanner graph plays more important roles for cycle
codes~\cite{ghol2,ghol3}, since it affects not only the message
dependence of iterative decoding but also the minimum distance of
the code~\cite{cycle2}.
\medskip\medskip

Accordingly we propose  a new method to construct mother matrices in order to construction of cycle codes with high girts and various lengths. 
\section{Preliminaries and Notations}
A {\it graph} $G$ is a triple consisting of a {\it vertex set}
$V(G)$, an {\it edge set} $E(G)$, and a relation of incidence that associates with each
edge two vertices (not necessarily distinct), called its {\it
endpoints}.

A {\it bipartite graph} is a graph $G=(V,E)$ that $V$ can be
divided into two disjoint sets $A$ and $B$, such that every edge
$e\in E$ connects a vertex in $A$ to one in $B$.
\medskip

A length-$l$ {\it walk} is a successive series of edges $e_i$ and vertices
$v_j$, such as $v_1$$e_1$$ v_2$$ e_2$$ \cdots$ $ v_l$$ e_l$$
v_{l+1}$, forming a continuous curve, i.e. each $e_i$ connects
$v_i$ to $v_{i+1}$. A walk is closed if
the initial and terminal vertices be same, i.e $v_1=v_{l+1}$.
A closed walk where the only end vertices are the same is a {\it
cycle}, i.e. $v_{i}\ne v_{j}$ for each $1\le i< j\le l+1$ except
$i=1$, $j={l+1}$. Also the {\it girth} of a graph can be defined
as the length of the shortest cycle.
\medskip

To the LDPC code with the parity-check matrix $H$, we can
associate a graph referred to as its {\it Tanner graph} (TG), which is
a bipartite graph where the two disjoint sets collect
the check nodes and the bit nodes associated to the rows and
columns of $H$, respectively. An edge connects a check node to a
bit node if a nonzero entry exists in the intersection of the
corresponding row and column of $H$.
\medskip

Let $m,s$ be nonnegative integers with $0\le s\le m-1$. The
$m\times m$ circulant permutation matrix shifted by $s$, ${\cal
I}_m^s$, is the matrix obtained from $m\times m$ identity matrix
${\cal I}_m$ by shifting rows $s$ positions to the bottom, that is
${\cal I}_m^s=(e_{i,j})_{m\times m}$ where
$e_{i,j}=1$, if $i-j=s\bmod m$; and $e_{i,j}=0$, otherwise. It is clear that ${\cal I}_m^0={\cal
I}$. For simplicity, ${\cal I}_m^s$ is denoted by ${\cal I}^s$
when $m$ is known.

The definition of block-structure graph, given in \cite{ghol5}, is as the following.
\subsection{Block-Structure Graph}
 Let $m$, $b$ and $\gamma$ be some positive integers and
$b<\gamma$. Let ${\cal H}=(H_{i,j})_{b\times \gamma}$, where each
$H_{i,j}$ is a $m\times m$ circulant permutation matrix or the
$m\times m$ zero matrix. Considering $\cal H$ as a ${b\times \gamma}$
matrix with $m\times m$ entries, we refer to $\cal H$ as a matrix having $b$ block-rows and $\gamma$ block-columns.
For simplicity, the matrix $\cal H$ and
the quasi-cyclic LDPC code with the parity-check matrix $\cal H$
are called as $m-$circulant matrix and $m-$circulant code,
respectively. Then we can define the block-structure graph
associated to $\cal H$, denoted by BSG$({\cal H})$, as the following:
\begin{defn}
\rm\label{def-bsg} Let $G$ be a graph with the vertex set
$V(G)=\{v_1,v_2,...,v_b\}$, where $v_i$ represents the $i$th
block-row of $\cal H$. For each $i,j\in \{1,\cdots,b\}$ and $k\in
\{1,\cdots,\gamma\}$, where $H_{i,k}={\cal I}^{s_1}$ and
$H_{j,k}={\cal I}^{s_2}$ for some $0\le s_1, s_2\le m-1$, two
vertices $v_i,v_j\in V$ are joined by two directed edges labeled
with $(k, s)$, from $v_i$ to $v_j$, and $(k, s')$, from $v_j$ to
$v_i$, where $s = -s'=s_2 - s_1 \bmod m$. For each edge of $G$,
the first and second component of its label $(k, s)$ are referred
to as the {\it column index} and {\it slope} of that edge,
respectively. The resulting graph $G$ is called the {\it
block-structure graph} of $H$, and is denoted by BSG$({\cal H})$.
\end{defn}
\begin{defn}\rm
Let $G$ be BSG$(\cal H)$, where $\cal H$ is a $m$-circulant matrix.
A length-$l$ closed walk in $G$ is given by a sequence of vertices
$v_{i_1},v_{i_2},\cdots,v_{i_{l}},v_{i_{l+1}}$, where
$i_{l+1}=i_1$, with edges $e_1,e_2,...,e_{l}$ such that for each
$1\le j \le l$ edge $e_j$ connect vertex $v_{i_j}$ to vertex
$v_{i_{j+1}}$ and if $(k_j,s_j)$ denotes the label $e_j$ from
$v_{i_j}$ to vertex $v_{i_{j+1}}$ then the following conditions are
hold:
\begin{enumerate}
\item Each edge $e_j$ in the sequence $e_1,e_2,...,e_{l}$
is repeated at most $m$ times; \item For each $1\le j \le l$,
$k_j\ne k_{j+1}$, where $k_{l+1}:=k_1$, i.e. the index columns of successive edges are different;  \item $\sum_{j=1}^{l}
s_j\equiv 0\pmod m$, i.e. the sum of slopes of edges are zero in modulus of $m$.
\end{enumerate}
\end{defn}
For simplicity, we can show this length-$l$ closed walk, or briefly $l$ closed walk, in $G$ with the following chain:
$$ \begin{array}{c}
v_{i_1}\stackrel{(k_1,s_1)}{\longrightarrow}
v_{i_2}\stackrel{(k_2,s_2)}{\longrightarrow} v_{i_3}\quad...\quad
v_{i_{l-1}}\stackrel{(k_{l-1},s_{l-1})}{\longrightarrow}v_{i_l}\stackrel{(k_l,s_l)}{\longrightarrow}v_{i_1}.\end{array}$$
%%%%%%%%%%%%%%%%%%%%%%%%%%%%%%%%%%%%%%%%%%%%%%%%%%%%%%%%%%%%%%%%%%%%%%%%%%%%%%%%%%%%%%%%%%%%%%%%%%%%%%%%%%%%%%%%%%%%%%%%%%%%%%%%%5
\begin{defn}\rm
Let $a,b$ and $c$ be some non-negative integers such that $a\ge2$, $b\ge1$ and $c\ge0$. By $(a,b,c)-$double-cylinder LDPC (DC-LDPC) codes, we mean QC LDPC codes having the mother matrix $H(a,b,c)$ as follows:
$$H(a,b,c)=\left(
    \begin{array}{cccccccccccccccc}
    \cline{1-4}
      \multicolumn{1}{|c}{1}&&\cdots&\multicolumn{1}{c|}{1}&&&&&&\cdots&&&&&\multicolumn{1}{|c|}{1}\\
      \cline{15-15}
      \multicolumn{1}{|c}{1} & 1 & & \multicolumn{1}{c|}{\vdots } &  &   &   &  \\
      \multicolumn{1}{|c}{} & \ddots &\ddots & \multicolumn{1}{c|}{\vdots } &   &   &   &  \\
      \cline{5-5}
        \multicolumn{1}{|c}{}&&1&\multicolumn{1}{c|}{1} & \multicolumn{1}{|c|}{1}  &   &   &   &   &  \\
        \cline{1-4}\cline{6-6}
        &&&&\multicolumn{1}{|c|}{1}&\multicolumn{1}{|c|}{1}\\
        \cline{5-5}
        &&&&&\multicolumn{1}{|c|}{1}\\ \cline{6-6}
        &&&&&&\ddots\\
        \cline{8-8}
        &&&&&&&\multicolumn{1}{|c|}{1}\\
        \cline{9-12}
        &&&&&&&\multicolumn{1}{|c|}{1}&\multicolumn{1}{|c}{1}&&\cdots&\multicolumn{1}{c|}{1}\\
        \cline{8-8}
        &&&&&&&&\multicolumn{1}{|c}{1} & 1 & & \multicolumn{1}{c|}{\vdots }\\
        &&&&&&&&\multicolumn{1}{|c}{} & \ddots &\ddots & \multicolumn{1}{c|}{\vdots} \\
        \cline{13-13}
        &&&&&&&&\multicolumn{1}{|c}{}&&1&\multicolumn{1}{c|}{1} & \multicolumn{1}{|c|}{1} \\
        \cline{9-12}
        &&&&&&&&&&&&\multicolumn{1}{|c|}{1}\\
        \cline{13-13}
        &&&&&&&&&&&&&\ddots\\
        \cline{15-15}
        &&&&&&&&&&&&&&\multicolumn{1}{|c|}{1}\\

    \end{array}
  \right)
$$
Where, the number of blocks
$\begin{tabular}{|c|}\hline% after \\: \hline or \cline{col1-col2} \cline{col3-col4} ...
1 \\1\\\hline\end{tabular}$
between each two consecutive blocks
$\begin{array}{|cccc|}
   \hline
   1 &  & \ldots & 1 \\
   1 & 1 &  & \vdots \\
    & \ddots & \ddots & \vdots \\
    &  & 1 & 1 \\
   \hline
 \end{array}_{\,\,a\times a}={\cal I}+{\cal I}^1$
is exactly $c$, and the number of blocks ${\cal I}+{\cal I}^1$ in $H(a,b,c)$ is $b$.
\end{defn}
\rm\label{bsg}
\begin{figure}
\centerline{\includegraphics[scale=0.8]{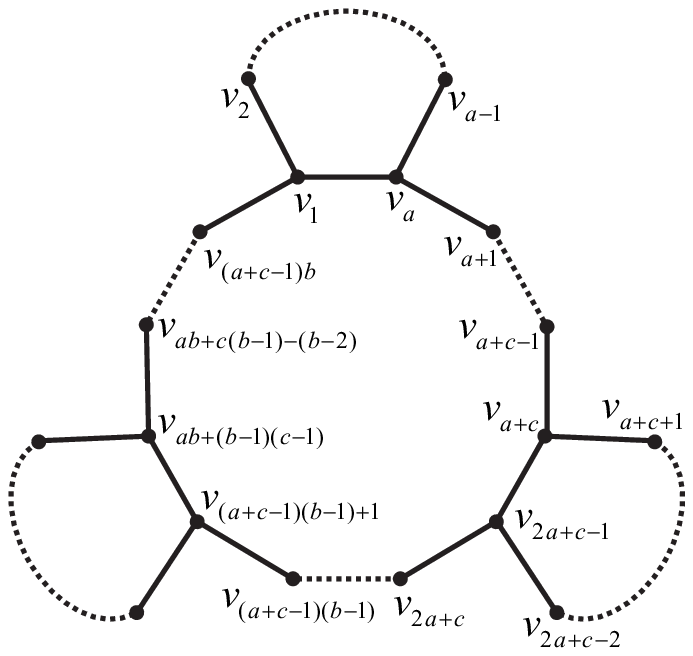}\hspace{0.8cm}
\includegraphics[scale=0.5]{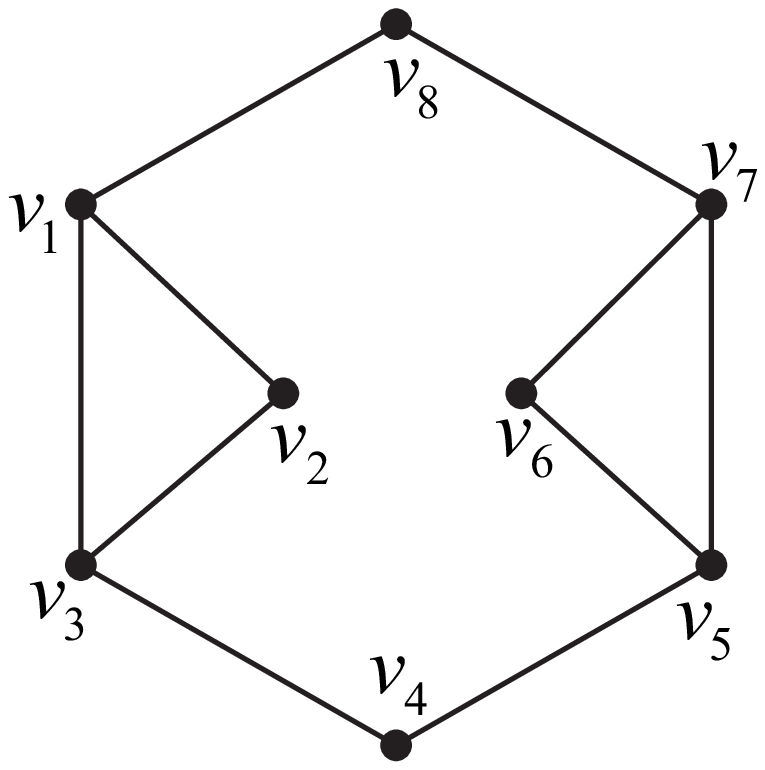}}
\caption{a) Overal diagram of {\rm BSG}$(H(a,b,c))$\quad b){\rm BSG}$(H(3,2,2))$ of Example~\ref{ex1}.}
\end{figure}

In fact, $H(a,b,c)=(h_{i,j})$ is a ${(a+c-1)b\times (a+c)b}$ binary
matrix, where $h_{i,j}=1$, if and only if one of two following
conditions is hold:
\begin{enumerate}
\item
$\lfloor\frac{i-1}{a+c-1}\rfloor=\lfloor\frac{j-1}{a+c}\rfloor$, and
if $i1=i-1\bmod{(a+c-1)}$, and $j1=j-1\bmod{(a+c)}$, then one of the
following conditions is hold:
\begin{enumerate}
\item $i1<a$, $j1<a$ and $i1-j1\stackrel{a}\equiv0,1$.
\item $i1\ge a-1$, $j1\ge a$ and $j1-i1=0,1$.
\end{enumerate}
\item $i=(a+c-1)*k+1\pmod{(a+c-1)*b}$, $j=(a+c)*k$, for some $1\le k\le b$.
\end{enumerate}
So, the design rate of the cycle code with the parity-check matrix $H(a,b,c)$ is ${\cal R}=1-\frac{(a+c-1)b}{(a+c)b}=\frac{1}{a+c}$; whereas the following lemma states the the maximum-achievable girth $g_{\max}$ of QC LDPC codes having the mother matrix $H(a,b,c)$, for $b$ large enough, is $8(a+c)$. This shows that larger rates yields more less maximum-achievable girths.
\begin{ex}\rm
Let $a=3$ and $b=c=2$. The parity-check matrix of $H(3,2,2)$ is

$$H(3,2,2)=\left(
    \begin{array}{cccccccccccc}
    \cline{1-3}
      \multicolumn{1}{|c}{1}&&\multicolumn{1}{c|}{1}&&&&&&&\multicolumn{1}{|c|}{1}\\
      \cline{10-10}
      \multicolumn{1}{|c}{1} & 1 &  \multicolumn{1}{c|}{} &   &   &   &  \\
      \cline{4-4}
        \multicolumn{1}{|c}{}&1&\multicolumn{1}{c|}{1} & \multicolumn{1}{|c|}{1}  &   &   &   &   &  \\
        \cline{1-3}\cline{5-5}
        &  &   & \multicolumn{1}{|c|}{1}  & \multicolumn{1}{|c|}{1} &   &   &   &   &  \\
        \cline{4-4}\cline{6-8}
        &   &   &   & \multicolumn{1}{|c|}{1} & \multicolumn{1}{|c}{1} &   & \multicolumn{1}{c|}{1} &   &  \\
        \cline{5-5}
        &   &   &   &   & \multicolumn{1}{|c}{1} & 1 & \multicolumn{1}{c|}{}  &   &  \\
        \cline{9-9}
        &   &   &   &   & \multicolumn{1}{|c}{}  & 1 & \multicolumn{1}{c|}{1} & \multicolumn{1}{|c|}{1}  &  \\
        \cline{6-8}\cline{10-10}
        &   &   &   &   &   &   &   & \multicolumn{1}{|c|}{1} & \multicolumn{1}{|c|}{1} \\
        \cline{9-9}
    \end{array}
  \right)
$$
\label{ex1}
\end{ex}
\begin{thm}\rm
Let $a\ge2$, $b\ge1$ and $c\ge0$ be some integers and $g_{\max}$ be the maximum achievable girth of QC LDPC codes having mother matrix $H(a,b,c)$.
If $b\ge \lceil\frac{a-1}{c+1}\rceil+2$, then $g_{\max}=8(c+a)$, else $g_{\max}=4(bc+b+a-1)$.

\pf
Let $H$ be the $(H(a,b,c))$, as shown in Figure~\ref{bsg}. Let ${\cal P}_1$ be the following chain from $v_a$ to itself:
$$v_{a}\stackrel{(a,s_a)}{\longrightarrow}v_{1}\stackrel{(1,s_1)}{\longrightarrow}v_{2}\stackrel{(2,s_2)}{\longrightarrow}v_3\cdots v_{a-1}\stackrel{(a-1,s_{a-1})}{\longrightarrow} v_{a},$$
${\cal P}_2$ be the following chain from $v_a$ to $v_{a+c}$:
$$v_{a}\stackrel{(a+1,s_{a+1})}{\longrightarrow}v_{a+1}\stackrel{(a+2,s_{a+2})}{\longrightarrow}v_{a+2}\cdots v_{a+c-1}\stackrel{(a+c-1,s_{a+c-1})}{\longrightarrow}v_{a+c},$$
and ${\cal P}_3$ be the following chain from $v_{a+c}$ to itself:
$$v_{a+c}\stackrel{(a+c,s_{a+c})}{\longrightarrow}v_{a+c+1}\cdots v_{2a+c-1}\stackrel{(2a+c-1,s_{2a+c-1})}{\longrightarrow}v_{a+c}.$$
It is clear that ${\cal P}_1$, ${\cal P}_2$ and ${\cal P}_3$ have length $a$, $c$ and $a$, respectively.
Hence, the chain ${\cal P}={\cal P}_1 {\cal P}_2 {\cal P}_3 {\cal P}_2^{-1} {\cal P}_1^{-1} {\cal P}_2 {\cal P}_3^{-1} {\cal P}_2^{-1}$ is an inevitable walk with length $l({\cal P})=2l({\cal P}_1)+4l({\cal P}_2)+2l({\cal P}_3)=4(a+c)$ from $v_a$ to itself, which is equivalent to an inevitable length-$8(a+c)$ in the Tanner graph. So, $g_{\max}\le 8(a+c)$. On the other hand, let ${\cal P}'_1$ be the chain from $v_1$ to $v_a$ as follows.
$$v_{1}\stackrel{(1,s_1)}{\longrightarrow}v_{2}\stackrel{(2,s_2)}{\longrightarrow}v_3\cdots v_{a-1}\stackrel{(a-1,s_{a-1})}{\longrightarrow} v_{a},$$
and ${\cal P}'_i$, $2\le i\le b+1$, be the following chains from $v_{(i-1)a+(i-2)(c-1)}$ to $v_{ia+(i-1)(c-1)}$:
$$v_{(i-1)a+(i-2)(c-1)}\stackrel{((i-1)a+(i-2)(c-1)+1,s_{(i-1)a+(i-2)(c-1)+1})}{\longrightarrow}v_{(i-1)a+(i-2)(c-1)+1}{\longrightarrow}\cdots\to v_{ia+(i-1)(c-1)},$$
where we accept that $v_{(b+1)a+b(c-1)-1}=v_1$ and $v_{(b+1)a+b(c-1)}=v_a$. It is clear that ${\cal P'}_1$ and ${\cal P'}_i$, $2\le i\le b+1$, have length $a-1$, $c+1$, respectively. Hence, the chain $${\cal P'}={\cal P'}_1 {\cal P'}_2 \ldots {\cal P'}_{b+1} {\cal P'}_1^{-1} {\cal P'}_{b+1}^{-1} \ldots {\cal P'}_2^{-1} $$ is an inevitable walk with length $l({\cal P'})=2\sum_{i=1}^{b+1}l({\cal P'}_i)=2((a-1)+b(c+1))=2(bc+b+a-1)$ from $v_a$ to itself, which is equivalent to an inevitable length-$4(bc+b+a-1)$ in the Tanner graph. So, $g_{\max}\le 4(bc+b+a-1)$. However $4(a+c)=l({\cal P})\ge l({\cal P'})=2(bc+b+a-1)$, if and and only if $b\ge \frac{a-1}{c+1}+2$, which is hold iff $b\ge \lceil\frac{a-1}{c+1}\rceil+2$.
\eep
\end{thm}
Figure \ref{abc} shows maximum-achievable girth for different a, b and c values. In the following we propose construction of QC-LDPC codes.
\label{abc}
\begin{figure}
\begin{center}
\centerline{\includegraphics[scale=0.6]{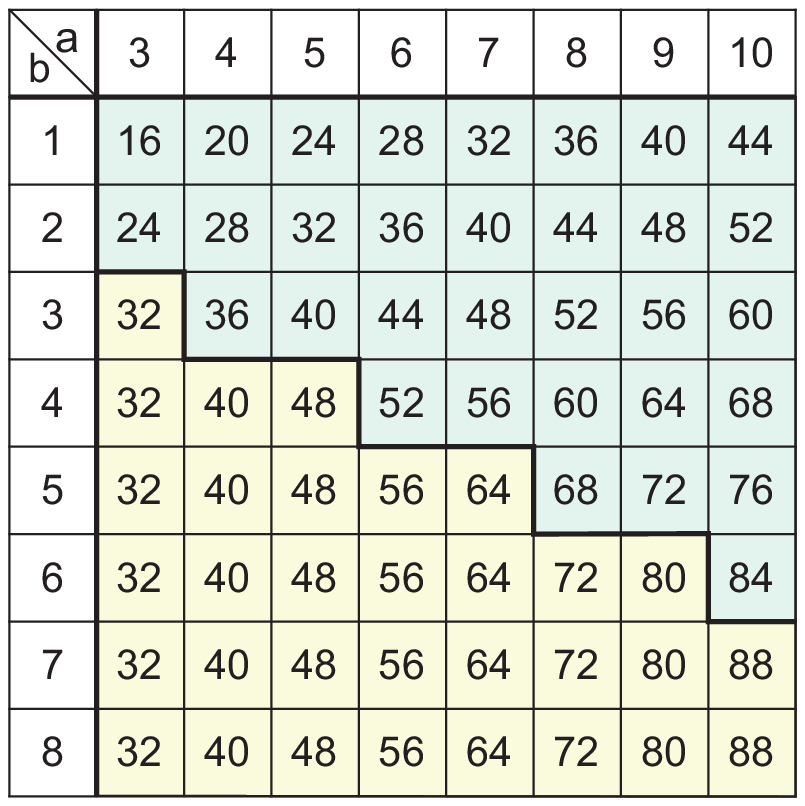}\hspace{0.5cm}
\includegraphics[scale=0.6]{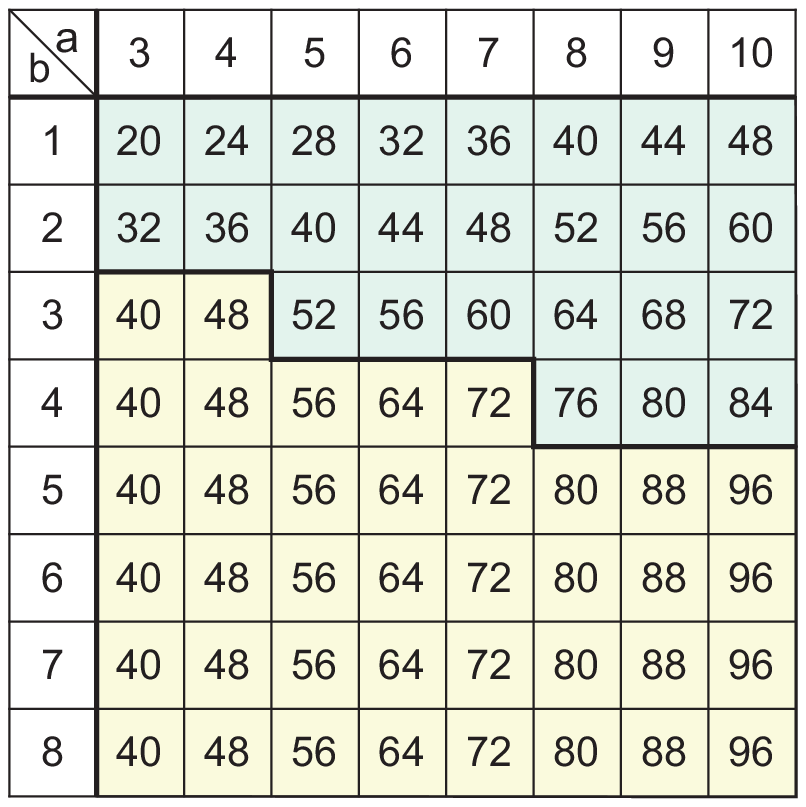}\hspace{0.5cm}\includegraphics[scale=0.6]{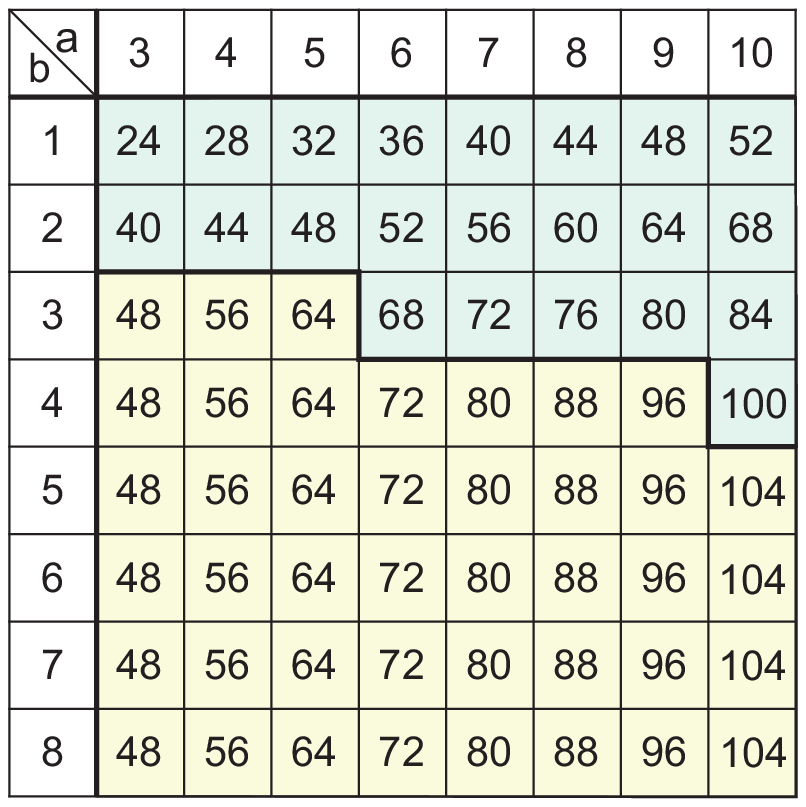}}
\end{center}
\caption{The maximum-achievable girth of $DC(a,b,c)$ codes for $c=1,2,3$, from left to right, respectively.}
\end{figure}
\subsection{Construction of DC-LDPC codes}
\label{secalg}
Given  positive integers $a$,$b$ and $c$, let $H(a,b,c)$ be the parity check matrix of a double-cylinder LDPC (DC-LDPC), by  code-generating algorithm given in~\cite{ghol4} we can obtain suitable shift sequence values by select an enough large $m$ as size of circulate permutation matrices  and in order to construct double-cylinder LDPC (DC-LDPC) code with desire girth. In table \ref{tab-code} some DC-LDPC codes various length $n$ rate $r$ and girth $2g$.
\medskip
%%%%%%%%%%%%%%%%%%%%%%%%%%%%%%%%%%%%%%%%%%
%%%%%%%%%%%%%%%%%%%%%%%%%%%%%%%%%%%%%%%%%%
%%%%%%%%%%%%%%%%%%%%%%%%5  Simulation result
%%%%%%%%%%%%%%%%%%%%%%%%%%%%%%%%%%%%%%%%%%%%%

\begin{table}[h]
\caption{FSS$(m, {\cal B}, S)$ codes with girth $2g$ and length $n$
constructed from FSS$(v, b, 2)$}
\tiny
$\hspace{-1cm}
\begin{array}{|c|c|c|c|c|c|c|p{14cm}|}
\hline \it  b&c&a&{\cal R}&g&m&n&S\\
\hline\hline
&&3&1/5&40&30&450&[0,28,19,5,16,14,25,10,15,16,13,4,6,3,25]\\
\cline{3-8}
&&4&1/6&48&42&756&[0,1,8,27,7,18,29,1,3,37,22,26,0,35,2,5,13,28]\\
\cline{3-8}
&&5&1/7&52&40&840&[0,1,9,22,19,6,6,6,4,2,0,16,5,1,33,0,3,27,28,26,38]\\
\cline{3-8}
&2&6&1/8&56&33&792&[0,0,0,0,0,1,0,0,0,0,0,0,0,3,0,0,0,0,0,0,0,5,0,21]\\
\cline{3-8}
&&7&1/9&60&28&756&[0,0,0,0,0,0,1,0,0,0,0,0,0,0,0,3,0,0,0,0,0,0,0,0,5,0,25]\\
\cline{3-8}
&&8&1/10&64&20&600&[0,0,0,0,0,0,0,1,0,0,0,0,0,0,0,0,0,3,0,0,0,0,0,0,0,1,3,0,0,13]\\
\cline{2-8}
&&3&1/6&48&45&810&[0,14,0,8,3,4,37,14,39,36,17,9,38,24,22,34,2,40]\\
\cline{3-8}
&3&4&1/7&56&30&630&[0,17,27,17,5,27,24,26,26,13,12,19,17,2,27,14,10,4,13,4,19]\\
\cline{3-8}
&&5&1/8&64&50&1200&[0,44,44,44,44,44,44,44,44,44,44,44,41,30,30,30,30,30,30,30,11,28,28,40]\\
\cline{2-8}
&&3&1/7&56&50&1050&[0,5,18,16,37,39,26,48,16,9,3,45,18,22,16,45,45,32,41,31,23]\\
\cline{3-8}
3&4&4&1/8&64&50&1200&[0,43,35,33,37,11,27,18,12,1,10,19,3,5,47,45,11,24,2,40,5,33,41,38]\\
\cline{3-8}
&&5&1/9&72&50 &1350&[0,10,36,47,12,39,2,11,4,7,25,28,27,48,21,45,10,0,7,22,5,9,14,33,18,20,12]\\
\cline{3-8}
&&6&1/10&80&50 &1500&[0,43,48,45,44,3,3,5,32,3,28,4,47,35,25,26,41,46,31,16,46,36,48,34,38,35,18,30,6,0]\\
\cline{2-8}
&&3&1/8&64&60 &1440&[0,9,46,35,15,25,56,22,15,42,46,41,18,3,42,57,6,11,18,38,53,14,54,0]\\
\cline{3-8}
&5&4&1/9&72&50 &1350&[0,46,1,15,25,35,0,10,32,36,10,5,0,11,15,17,13,19,33,7,1,26,10,34,44,10,40]\\
\cline{3-8}
&&5&1/10&80&50 &1500&[0,46,1,15,25,35,0,10,32,36,10,5,0,11,15,17,13,19,33,7,1,26,10,34,44,10,40]\\
\cline{2-8}
&6&3&1/9&72&60 &1650&[0,10,42,10,10,10,10,10,10,10,10,23,45,45,45,45,45,45,45,45,55,42,42,42,42,42,8]\\
\cline{3-8}
&&4&1/10&80&65&2100&[0,16,61,31,64,35,19,44,33,37,60,5,6,30,38,39,45,25,27,68,27,11,21,12,60,13,13,66,7,20]\\
\cline{1-8}
&&3&1/5&40&40&800&[0,5,2,24,18,37,27,3,6,21,38,29,32,32,28,26,16,29,24,31]\\
\cline{3-8}
&&4&1/6&48&48&1152&[0,27,31,21,4,22,23,19,25,6,2,10,20,30,25,22,9,30,16,8,28,17,38,1]\\
\cline{3-8}
&2&5&1/7&56&45&1260&[0,21,20,17,15,43,43,27,26,42,19,35,8,28,27,13,30,34,34,30,3,15,33,36,26,36,40,19]\\
\cline{3-8}
&&6&1/8&64&40&1280&[0,14,13,22,36,28,5,26,30,14,23,7,1,2,36,1,38,11,8,30,27,19,17,31,31,20,21,32,0,25,18,20]\\
\cline{3-8}
&&7&1/9&72&55&2420&[0,18,19,17,4,11,29,30,43,4,21,14,12,37,8,42,37,32,32,49,26,40,19,18,9,17,6,30,32,39,43,32,38,20,31,43]\\
\cline{3-8}
&&8&1/10&76&75&3000&[0,3,7,45,33,27,7,26,66,37,50,53,64,14,62,29,24,51,39,34,38,14,57,18,26,11,43,70,26,56,1,23,6,38,56,70,37,65,55,73]\\
\cline{2-8}
&&3&1/6&48&70&1680&[0,26,8,26,26,26,26,26,63,39,39,39,39,39,39,8,8,8,8,8,13,35,35,19]\\
\cline{3-8}
&3&4&1/7&56&50&1400&[0,32,27,37,16,33,33,3,24,20,42,19,44,26,19,22,32,35,25,4,45,36,45,0,32,21,15,46]\\
\cline{3-8}
&&5&1/8&64&40&1280&[0,0,2,20,11,13,24,2,19,0,31,22,29,36,26,33,23,30,21,33,14,19,35,1,15,31,27,25,23,38,1,13]\\
\cline{2-8}
4&&3&1/7&56&40&1120&[0,16,19,38,24,4,5,19,24,36,22,4,27,24,16,30,23,28,26,15,15,24,25,22,16,19,19,38]\\
\cline{3-8}
&4&4&1/8&64&55&1760&[0,25,37,3,42,17,31,0,39,0,32,43,20,26,31,29,4,11,14,34,37,21,7,35,0,35,47,14,6,3,25,18]\\
\cline{3-8}
&&5&1/9&72&55&1980&[0,52,8,26,0,7,34,2,4,1,30,19,25,43,46,24,15,29,30,49,3,47,44,25,19,16,32,4,52,33,17,14,31,4,31,32]\\
\cline{3-8}
&&6&1/10&80&60&2160&[0,46,9,12,13,34,35,53,55,55,5,41,13,22,44,39,2,20,43,8,29,31,50,57,52,58,54,1,45,36,17,47,32,47,41,35,22,7,6,43]\\
\cline{2-8}
&&3&1/8&64&40&1440&[0,31,4,31,37,29,12,7,31,18,20,0,5,10,15,26,37,22,0,38,33,14,25,15,29,27,19,28,15,25,21,33]\\
\cline{3-8}
&5&4&1/9&72&50&1600&[0,30,26,35,21,37,15,32,42,47,39,19,29,20,27,37,14,6,42,4,22,1,5,33,20,19,13,13,17,21,32,15,25,32,39,38]\\
\cline{3-8}
&&5&1/10&80&65&2600&[0,20,20,20,20,20,20,20,20,20,20,20,20,20,24,55,55,55,55,55,55,55,55,55,55,39,39,39,39,39,39,39,39,39,48,30,30,30,20,24]\\
\cline{2-8}
&6&3&1/9&72&40&1440&[0,28,25,25,0,32,21,15,8,23,10,4,5,17,24,29,0,24,0,26,7,22,33,25,37,29,37,10,38,35,2,17,22,11,32,30]\\
\cline{3-8}
&&4&1/10&80&40&1600&[0,18,33,10,24,13,35,35,11,18,27,21,24,15,26,32,38,3,18,17,28,35,0,16,29,36,6,24,10,38,15,16,4,4,14,3,32,38,7,4]\\
\cline{2-8}
&7&3&1/10&80&45&1800&[0,38,30,35,38,6,38,43,29,3,8,36,43,5,26,30,33,35,27,15,17,10,8,15,27,9,20,25,26,1,32,43,2,28,38,34,24,40,36,9]\\
\cline{1-8}
&&3&1/5&40&30&750&[0,25,26,0,12,13,11,13,24,25,6,3,16,15,17,10,11,24,23,5,9,11,7,8,0]\\
\cline{3-8}
&&4&1/6&48&35&1050&[0,26,15,1,2,5,12,22,11,22,21,0,4,25,13,5,23,1,24,25,28,26,31,13,25,15,5,1,27,25]\\
\cline{3-8}
&&5&1/7&56&45&1575&[0,41,13,20,6,23,28,8,39,26,30,42,39,37,4,14,19,9,6,42,31,23,41,14,15,20,14,38,3,6,20,1,26,3,25]\\
\cline{3-8}
&2&6&1/8&64&50&2000&[0,35,30,25,17,38,33,31,9,45,24,31,41,39,1,22,21,23,15,27,8,47,24,31,36,45,38,30,12,28,39,15,1,41,8,8,14,27,25,18]\\
\cline{3-8}
&&7&1/9&72&55&2475&[0,28,32,31,22,9,38,8,29,50,16,28,4,49,45,33,53,41,25,12,45,0,21,32,6,14,8,12,8,26,20,35,30,42,22,16,14,23,20,21,2,34,50,22,30]\\
\cline{3-8}
&&8&1/10&80&60&3500&[0,1,39,23,37,37,44,13,22,56,1,27,32,6,56,31,33,29,44,38,8,1,2,34,24,1,11,24,9,0,20,39,58,6,1,22,39,57,36,41,50,3,54,41,23,58,48, 5,50,54]\\
\cline{2-8}
&&3&1/6&48&30&900&[0,18,11,1,19,4,7,4,24,10,3,7,2,12,13,21,28,19,20,14,13,26,2,24,24,21,26,11,26,22]\\
\cline{3-8}
&3&4&1/7&56&40&1400&[0,23,28,21,22,36,12,12,11,13,6,26,27,28,4,12,22,10,20,4,25,19,26,33,25,35,27,6,10,19]\\
\cline{3-8}
&&5&1/8&64&45&1800&[0,34,2,15,4,13,33,30,23,23,30,0,7,24,23,25,11,6,21,32,42,26,0,36,6,0,41,35,24,7,3,14,39,36,8,31,19,35,14,18]\\
\cline{2-8}
5&&4&1/8&64&25&1000&[0,2,20,15,9,2,6,22,7,2,6,7,4,14,12,9,8,1,15,22,20,6,23,8,21,3,11,23,17,3,10,10,7,2,22,9,10,3,13,11]\\
\cline{3-8}
&4&5&1/9&72&45&2025&[0,42,19,0,18,9,21,24,25,0,41,25,28,21,18,39,22,22,8,37,21,43,40,5,9,20,39,21,25,40,39,40,43,38,19,19,2,30,0,31,5,23,23,7,34]\\
\cline{3-8}
&&6&1/10&80&50&2500&[0,21,52,36,31,53,37,18,15,14,21,16,29,38,6,31,41,40,1,13,46,6,23,23,14,35,2,24,52,1,9,24,32,46,13,43,52,54,13,2,58,58,56,5,56, 18,55,26,50,2]\\
\cline{2-8}
&&3&1/8&64&30&1600&[0,1,28,22,7,28,22,22,26,1,6,11,22,35,12,12,10,20,21,15,17,19,4,9,10,4,15,20,20,33,34,22,29,13,25,38,33,35,12,26]\\
\cline{3-8}
&5&4&1/9&72&40&2250&[0,32,32,19,9,20,18,24,37,18,14,18,27,38,26,25,15,16,43,9,48,36,45,6,47,11,37,47,22,47,33,9,40,27,3,34,8,31,36,38,25,44,7,44,3]\\
\cline{3-8}
&&5&1/10&80&50&2750&[0,5,38,4,34,29,20,32,6,20,24,11,14,34,24,37,34,43,34,27,40,46,35,14,15,10,8,45,10,25,23,28,35,22,15,26,39,37,23,30,37,1,34,13, 44,24,3,34,40,0,6,19,42,19,12]\\
%\cline{2-8}
%&6&3&1/9&64&35&1575&[0,5,38,4,34,29,20,32,6,20,24,11,14,34,24,37,34,43,34,27,40,46,35,14,15,10,8,45,10,25,23,28,35,22,15,26,39,37,23,30,37,1,34,13, 44,24,3,34,40,0,6,19,42,19,12]\\
%\cline{3-8}
%&&4&1/10&72&30&1500&[0,6,19,14,6,8,26,9,6,6,6,14,17,8,26,17,28,27,19,11,2,10,12,16,9,6,20,2,17,12,25,16,10,12,7,18,15,19,2,10,11,15,0,24,11,3,5,9,24,7]\\
%\cline{2-8}
%&7&3&1/10&72&40&2000&[0,8,37,16,0,23,19,31,23,33,36,4,3,35,17,26,9,9,9,20,30,2,25,38,19,15,8,23,34,11,31,13,13,19,4,30,22,3,24,23,11,32,2,12,26,2,5,16, 9,21]\\
\cline{1-8}
\end{array}
\label{tab-code}$
\end{table}


\begin{thebibliography}{99}
%%%%%%%%%%%%%%%%%%%%%%%%%%%%%%%%%%%%%%%%%%%%%%%%%%%%%%%%%%%%%%16
\bibitem{ghol5} {M. Gholami, and M. Esmaeili}, \emph{Maximum-girth Cylinder-type Block-circulant LDPC
Codes}, IEEE Trans. on Commun., vol. 60, no. 4, pp. 952--962, April 2012.
%%%%%%%%%%%%%%%%%%%%%%%%%%%%%%%%%%%%%%%%%%%%% 1  %%%%%%%%%%%%%%%%%%%%%%%%%%%%%%%%%%%%%%%%%%%%%%%%
\bibitem{gal} {R. G. Gallager}, ``Low-density parity-check codes," IRE Trans. Inf.
Theory, vol. IT-8, no. 1, pp. 21–-28, Jan. 1962. D. J. C.
%%%%%%%%%%%%%%%%%%%%%%%%%%%%%%%%%%%%%%%%%%%%% 2  %%%%%%%%%%%%%%%%%%%%%%%%%%%%%%%%%%%%%%%%%%%%%%%%
\bibitem{mac1} {D. J. C. MacKey}, ``Good error-correcting codes based on very sparse matrices," IEEE Trans.
Inf. Theory, vol. 45, no. 2, pp. 399–432, Mar. 1999.

%%%%%%%%%%%%%%%%%%%%%%%%%%%%%%%%%%%%%%%%%%%%% 3  %%%%%%%%%%%%%%%%%%%%%%%%%%%%%%%%%%%%%%%
\bibitem{richard} {S. Y. Chung, G. D. Forney, T. J. Richardson, and R. L.
Urbanke}, ``On the design of low-density parity-check codes within
0.0045 dB of the Shannon limit," IEEE Commun. Lett., vol. 5, no.
2, pp. 58–60, Feb. 2001.

%%%%%%%%%%%%%%%%%%%%%%%%%%%%%%%%%%%%%%%%%%%%% 4  %%%%%%%%%%%%%%%%%%%%%%%%%%%%%%%%%%%%%%%%%%%%%%%%
\bibitem{sum} {F. R. Kschischang, B. J. Frey and H. -A. Loeliger,} ``Factor graphs and the sum-product algorithm," IEEE
TIT: IEEE Trans. on Inf. Theory, vol 47, 2001.

%%%%%%%%%%%%%%%%%%%%%%%%%%%%%%%%%%%%%%%%%%%%% 5  %%%%%%%%%%%%%%%%%%%%%%%%%%%%%%%%%%%%%%%%%%%%%%%%
\bibitem{bone} {N. Bonello, S. Chen, and L. Hanzo}, \emph{Design of low-density parity-check
codes}, an overview, IEEE Vehicular Technology Magazine, Dec.
2011.

%%%%%%%%%%%%%%%%%%%%%%%%%%%%%%%%%%%%%%%%%%%%%%%%%%%%%%%%%%%%%%% 6 %%%%%%%%%%%%%%%%%%%%%%%%%%%%%%%%%%%%%%%%%%%%%%%%%%%%%%%%%%%%%%%%%%%
\bibitem{tan1} {R.M. Tanner, D. Sridhara, A. Sridharan, T.E. Fuja, and D.J. Costello}, ``LDPC
block and convolutional codes based on circulant matrices," IEEE
Trans. Inform. Theory, vol. 50, no. 12, pp. 2966-–2984, Dec. 2004.
%%%%%%%%%%%%%%%%%%%%%%%%%%%%%%%%%%%%%%%%%%%%%%%%%%%%%%%%%%%%%%%%%%%%%% 7
\bibitem{sun} {Sunghwan Kim, Jong-Seon No, Habong Chung, and Dong-Joon Shin}, \emph{Quasi-cyclic low-
density parity-check codes with girth larger than 12}, IEEE Trans.
Inform. Theory, vol. 53, no. 8, pp. 2885--2891, Aug. 2007.
%%%%%%%%%%%%%%%%%%%%%%%%%%%%%%%%%%%%%%%%%%%%% 8 %%%%%%%%%%%%%%%%%%%%%%%%%%%%%%%%%%%%%%%
\bibitem{mac2} {M. C. Davey and D. J. C. MacKay}, ``Low-density parity check
codes over GF(q)," IEEE Commun. Lett., vol. 2, no. 6, pp.
165--167, June 1998.
%%%%%%%%%%%%%%%%%%%%%%%%%%%%%%%%%%%%%%%%%%%%%%%%%%%%%%%%%%%%%%%%%%%%%%% 9
\bibitem{cyclecodes} {T. D. Souza}, \emph{Cycle Codes}, EPFL / ALGO-LMA, SSC 6th semester, June. 2005.

%%%%%%%%%%%%%%%%%%%%%%%%%%%%%%%%%%%%%%%%%%%%%%%%%%%%%%%%%%%%%%%%% 10 %%%%%%%%%%%%%%%%%%%%%%%%%%%%%%%%%%%%%%%%%%%%%%%%%%%%%%%%%%%%%%
\bibitem{song1} {H. Song, J. Liu, B. V. K. V. Kumar}, ``Low complexity LDPC codes for
magnetic recording," in IEEE Globecom 2002, Taipei, Taiwan,
R.O.C., Nov. 2002.

%%%%%%%%%%%%%%%%%%%%%%%%%%%%%%%%%%%%%%%%%%%%%%%%%%%%%%%%%%%%%%%%% 11 %%%%%%%%%%%%%%%%%%%%%%%%%%%%%%%%%%%%%%%%%%%%%%%%%%%%%%%%%%%%%%
\bibitem{song3} {H. Song, J. Liu, B. V. K. V. Kumar}, ``Large girth cycle codes for partial response
channels," IEEE Trans. Magn., vol. 40, no. 4, part 2, pp.
3084–-3086, 2004.

%%%%%%%%%%%%%%%%%%%%%%%%%%%%%%%%%%%%%%%%%%%%%%%%%%%%%%%%%%%%%%%%% 12 %%%%%%%%%%%%%%%%%%%%%%%%%%%%%%%%%%%%%%%%%%%%%%%%%%%%%%%%%%%%%%
\bibitem{cycle1} {X.-Y. Hu, and E. Eleftheriou}, ``Binary representation of cycle
Tannergraph codes," in Proc. IEEE Intern. Conf. on Commun., Paris,
France, pp. 528--532, June 2004.
%%%%%%%%%%%%%%%%%%%%%%%%%%%%%%%%%%%%%%%%%%%%%%%%%%%%%%%%%%%%%%%%% 13 %%%%%%%%%%%%%%%%%%%%%%%%%%%%%%%%%%%%%%%%%%%%%%%%%%%%%%%%%%%%%%
\bibitem{cycle2} {C. Poulliat, M. Fossorier, and D. Declercq}, ``Design of regular
$(2, d_c)$ LDPC codes over GF(q) using their binary images," IEEE
Trans. Commun., vol. 56, pp. 1626--1635, Oct. 2008.

%%%%%%%%%%%%%%%%%%%%%%%%%%%%%%%%%%%%%%%%%%%%%%%%%%%%%%%%%%%%%%%%% 14 %%%%%%%%%%%%%%%%%%%%%%%%%%%%%%%%%%%%%%%%%%%%%%%%%%%%%%%%%%%%%%
\bibitem{ghol2} {M. Esmaeili, and M. Gholami}, \emph{Geometrically-structured maximum-girth LDPC block and
convolutional codes}, IEEE journal of selected area of commun., vol. 27, no. 6, Aug. 2009.
%%%%%%%%%%%%%%%%%%%%%%%%%%%%%%%%%%%%%%%%%%%%%%%%%%%%%%%%%%%%%%%%% 15 %%%%%%%%%%%%%%%%%%%%%%%%%%%%%%%%%%%%%%%%%%%%%%%%%%%%%%%%%%%%%%
\bibitem{ghol3} {M. Esmaeili, and M. Gholami}, \emph{Maximum-girth Slope-based Quasi-cyclic $(2,k\geq 5)$-LDPC  Codes}, IET commun., vol. 2, no. 10, Aug. 2008.


\end{thebibliography}
\end{document}